\documentclass[1p]{elsarticle}  

\usepackage{bm,bbm}     
\usepackage{amssymb,amsmath}   
\usepackage{slashed}  
\usepackage{natbib}

\newcommand{\pmu}{\partial_{\mu} }
\newcommand{\pnu}{\partial_{\nu} }
\newcommand{\pMu}{\partial^{\mu} }
\newcommand{\pNu}{\partial^{\nu} }
\newcommand{\Pmu}{\nabla_{\mu} }
\newcommand{\Pnu}{\nabla_{\nu} }
\newcommand{\PMu}{\nabla^{\mu} }
\newcommand{\PNu}{\nabla^{\nu} }

\newcommand{\tDelta}{\tilde{\Delta}}
\newcommand{\ee}{\mathrm{e}}

\newcommand{\eab}{\epsilon_{ab}}
\newcommand{\emn}{\epsilon^{\mu\nu}}
\newcommand{\ii}{\text{i}}
\newcommand{\id}{\mathbbm{1}}

\begin{document}

\title{Renormalization of the Nonlinear ${\rm O}(3)$ Model with $\theta$-Term}

\author{Raphael Flore}
\ead{raphael.flore@uni-jena.de}
\address{Theoretisch-Physikalisches Institut, Friedrich-Schiller-Universit\"at Jena, Max-Wien-Platz 1, D-07743 Jena, Germany \\ (Phone: +49 3641 947104, Fax: +49 3641 947132)}

\begin{abstract}
The renormalization of the topological term in the two-dimensional nonlinear O$(3)$ model is studied by means of the Functional Renormalization Group. By considering the topological charge as a limit of a more general operator, it is shown that a finite multiplicative renormalization occurs in the extreme infrared. In order to compute the effects of the zero modes, a specific representation of the Clifford algebra is developed which allows to reformulate the bosonic problem in terms of Dirac operators and to employ the index theorem.
\end{abstract}


\maketitle

\section{Introduction}\label{introduction}
One of the most interesting characteristics of the two-dimensional O$(3) \cong CP^1$ model is the nontrivial topology of the target space which allows for instantons and the definition of a topological charge $Q$ which represents the winding number of the field configurations. The inclusion of this topological charge as $\theta$-term in the action has attracted much attention since Haldane showed that antiferromagnetic spin-$S$ chains can be mapped onto the O$(3)$ model with $\theta = 2\pi S$ \cite{Haldane}. The physical properties of the model depend nontrivially on the topological parameter, most prominently the mass gap which vanishes for $\theta= \pi$ \cite{Shankar}. Furthermore, the vacuum energy density is a function of $\theta$ which can be seen in a large-$N$ expansion as well as a dilute instanton gas approximation, cf. \cite{Asorey} and references therein. This $\theta$-dependence of the mass gap and vacuum energy are also confirmed by numerical simulations, see e.g. \cite{Alles} and \cite{Boegli}. More information about lattice computations of the sigma model with topological term are given in \cite{Vicari}. More recently, the case $\theta$ slightly below $\pi$ was considered as a toy model for walking technicolor \cite{Pepe,Nogradi}.\\
Since the winding number is not altered by fluctuations, one would naively expect that this topological operator is not renormalized. In addition, it was explicitly shown in \cite{Boegli,Nogradi} that since the topological charge distinguishes between different vacua, it cannot be an irrelevant operator that renormalizes to zero. On the other hand, the investigation of non-Abelian gauge theories, which share interesting properties with the sigma model, indicated that a finite renormalization of the $\theta$-parameter occurs in the extreme momentum ranges. These nontrivial effects were first studied in \cite{Anselm,Shifman} and \cite{Johansen}, and subsequently also by means of the Functional Renormalization Group (FRG) \cite{ReuterTopo}. The result of the latter investigation was a finite, discrete renormalization of $\theta$ in the extreme UV and the extreme IR. A similar behavior in the extreme IR was found in an analysis of the coupling of Chern-Simons theory \cite{Chern}.\\ 
The purpose of this article is to investigate if a similar renormalization of the topological parameter also occurs in nonlinear sigma models. The FRG formalism is an appropriate framework to address this question. We will follow \cite{ReuterTopo} and analyze a more general class of operators by considering a spacetime-dependent coupling $\theta \to \theta \alpha(x)$, where $\alpha$ is an auxiliary scalar field. At the end we will set $\alpha(x)\to 1$. The problem shall be studied in Euclidean spacetime and we consider the action in the covariant formulation
\begin{align}
\label{classicAnsatz}
S_{\theta} ~ = S + \ii \,\theta Q ~ =& \,\frac{1}{2}\, \zeta \int d^2 x ~ h_{ab}(\phi) \pmu \phi^a \partial^{\mu} \phi^b \,+ \frac{\ii}{2\pi} \theta \int d^2 x \,\emn \sqrt{h} \eab ~\alpha~   \pmu \phi^a \pnu \phi^b\,, 
\end{align}
where the fields are maps $\phi : \mathbb{R}^2 \rightarrow S^2$, $h_{ab}(\phi)$ is the metric on the sphere and $h$ its determinant.\\
The article is structured as follows: A covariant formulation of the flow equation of the model is derived in Sec.\,2, before the renormalization of the coupling $\zeta$ is discussed in Sec. 3. Thereafter the renormalization of $\theta$ is analyzed, first in the UV  (Sec. 4) and then in the IR (Sec. 5). Finally, the conclusions are presented in Sec. 6.

\section{The Functional RG of the Model}

The Functional Renormalization Group describes the renormalization of a theory by means of a flow equation for the Effective Average Action $\Gamma_k$ which depends on the momentum scale $k$ and interpolates between the bare action at the scale $k\rightarrow \infty$ and the full effective action at $k=0$ \cite{FRG}. The interpolation of $\Gamma_k$ is described by a flow equation, as it is given in Eq. \eqref{erge2}, which is an exact relation and includes all orders of perturbation theory. However, in explicit computations it is in general impossible to take all terms into account that are generated in the effective action. One has to approximate the computation by truncating $\Gamma_k$ to a finite number of operators. Assuming that the operators of the bare action are the dominant ones, the ansatz 
\begin{align}
\label{topoAnsatz}
\Gamma_k[\phi] ~ = &\,\frac{1}{2} \zeta_k \int d^2 x ~ h_{ab}(\phi) \pmu \phi^a \partial^{\mu} \phi^b \,+ \frac{\ii}{2\pi} \theta_k \int d^2 x \emn \sqrt{h} \eab ~\alpha~   \pmu \phi^a \pnu \phi^b\,,
\end{align}
will be studied. Note that, different to \eqref{classicAnsatz}, $\phi$ denotes average fields and the couplings are not the bare ones but running. The flow equation depends on the second variation of the action functional and in order to obtain a covariant formulation of this expression, the background field expansion suggested in \cite{Alvarez,mukhi} will be utilized. This expansion and its application within the FRG framework shall be depicted here only briefly, while a more detailed discussion is given in \cite{PercOmar,FRGeigen}.\\
If $\varphi$ denotes the background field and $\phi$ is sufficiently close to $\varphi$, there is a unique geodesic connecting both fields and one can construct the ``exponential map''
\begin{equation}\label{exponential map}
 \phi^a = {\rm Exp}_{\varphi} \xi^a = \phi^a(\varphi,\xi)\,
\end{equation}
in which $\xi^a$ is the tangent vector to the geodesic at $\varphi$. This geodesic can be parametrized by an affine parameter $\lambda\in [0,1]$ as $\varphi_\lambda$ such that $\varphi_0=\varphi$ and $\varphi_1=\phi$. The tangent vector at a generic point $\varphi_\lambda$ is denoted by $\xi_\lambda = d\varphi_\lambda/d\lambda$. By means of the derivative along the geodesic, $\nabla_\lambda\equiv \xi^a_\lambda\nabla_a$, the covariant background field expansion takes the form \cite{mukhi}:
\begin{equation}\label{expansion2}
\Gamma_k[\phi] = \Gamma_k[\varphi_\lambda]\big|_{\lambda=1}
 = \sum_{n \ge 0} \left.\frac{1}{n!} \frac{d^n}{d\lambda^n}  \Gamma_k[\varphi_\lambda]\right|_{\lambda=0}
 = \sum_{n \ge 0} \left.\frac{1}{n!} \nabla_\lambda^n  \Gamma_k[\varphi_\lambda]\right|_{\lambda=0}\nonumber
\end{equation}
The expansion of the ansatz \eqref{topoAnsatz} reads
\begin{align}\label{expansion}
 & \Gamma_k[\phi] = \Gamma_k[\varphi,\xi] \\
&= \Gamma_k[\varphi] ~ + \frac{\zeta_k}{2}  \int \! d^dx ~ 2 h_{ab}\partial_\mu \varphi^a\nabla^\mu \xi^b + \nabla_\mu \xi^a\nabla^\mu \xi_a + R_{abcd}\partial_\mu\varphi^b\partial^\mu\varphi^c\xi^a\xi^d \nonumber  \\
 &~~~ + \frac{\ii \theta_k}{2\pi} \int d^2x \emn \sqrt{h} \eab ~ \alpha \left(2 \partial_\mu \varphi^a \Pnu \xi^b + \Pmu \xi^a \Pnu \xi^b + {R^a}_{cde} \pmu \varphi^e \pnu \varphi^b \xi^c \xi^d \right) + {\cal O}(\xi^3)\,.\nonumber
\end{align}
The covariant spacetime derivative of the pullback of a vector $\xi^a$ is defined as $\nabla_\mu \xi^a \equiv \partial_\mu \xi^a + \Gamma^a{}_b{}_c \, \partial_\mu \varphi^b\,\xi^c$. The regularization of the path integral is given in the FRG approach by the introduction of a cutoff action $\Delta S_k$. It is chosen such that the infrared contributions of the fluctuations $\xi$ below the scale $k$ are suppressed while the modes above $k$ are integrated out, providing in this way the correct interpolation of $\Gamma_k$. An appropriate form is
\begin{align}\label{cutoff}
 &\Delta S_k[\varphi,\xi] = \frac{1}{2} \int d^dx ~ \xi^a {\cal R}^k_{ab}(\varphi) \xi^b\,.
\end{align}
$\text{with}~\lim_{k\to 0}{\cal R}^{k}_{ab}[\varphi]=0~~\text{and}~\lim_{k\rightarrow \infty}{\cal R}^{k}_{ab}[\varphi]=\infty$. The flow equation for the Effective Average Action is then given as \cite{FRG}:
\begin{equation}\label{erge2}
k\partial_k \Gamma_k[\varphi,\xi] = \frac{1}{2}{\rm Tr}\left(\frac{k\partial_k{\cal R}_k[\varphi]}{\Gamma_k^{(0,2)}[\varphi,\xi]+{\cal R}_k[\varphi]}\right)\,.
\end{equation}
The equation has a concise one-loop structure, but it is important to note that it is defined in terms of the full propagator and, for the full $\Gamma_k$, includes all loop orders of perturbation theory. The usual one-loop result could be obtained if one inserted the second derivative of the bare action $S$ instead of $\Gamma_k$ into the denominator on the r.h.s. of \eqref{erge2}. The simple ansatz \eqref{topoAnsatz} studied in this paper only takes care of the renormalization of the operators that belong to the bare action, but it accounts for the running of the parameters and hence corresponds to an RG-improved one-loop calculation\footnote{Note that the approximation scheme of the FRG, which is based on truncations of the effective action, is in general completely different from the approximation scheme of perturbation theory and not directly comparable for higher loop orders.}.\\
In the following we will utilize the notation $k\partial_k A_k = \dot{A}_k$ for any $k$-dependent object $A_k$. The flow equation holds true independent of the specific field configuration and we will evaluate it at\footnote{Apart from being a convenient choice, it is the only way to construct an effective action which is a functional of a single field only instead of $\varphi$ and $\xi$ separately. Cf. \cite{FRGeigen} for a more detailed discussion.} $\xi=0$, i.e. $\phi=\varphi$. With the commutator of the covariant spacetime derivatives, $H^{\mu \nu}_{ab} \equiv [\PMu,\PNu]_{ab} = R_{abcd} \pMu \varphi^c \pNu \varphi^d$, and the first Bianchi identity, one can compute $\Gamma_{k,ab}^{(0,2)}[\varphi,0]$ from Eq. \eqref{expansion} as
\begin{align}
\label{variation}
 \Gamma^{(0,2)}_{k,ab}[\varphi,0] &= - \zeta_k (\PMu\Pmu)_{ab} + \zeta_k \underbrace{R_{acdb} \pmu \varphi^c \partial^{\mu} \varphi^d}_{=M_{ab}} - \underbrace{ \frac{\ii}{\pi} \theta_k \, \emn (\pmu \alpha) \sqrt{h} \epsilon_{ac} \, {\Pnu^c}_{,b} }_{= B_{ab}(\alpha)} \nonumber\\
 &\equiv \zeta_k \tDelta_{ab} - B_{ab}(\alpha)\,,
\end{align}
where $\tDelta$ denotes the Laplacian operator $\tDelta_{ab} \equiv -(\Pmu\PMu)_{ab} + M_{ab}$. The result in \eqref{variation} shows that the flow equation is sensitive to the topological term, only if it is considered in a generalized form which contains a spacetime-dependent auxiliary field..\\
As the physical properties of the system should be independent of the specific regularization scheme, there is some freedom to choose an appropriate ``regulator'' $\mathcal{R}_k$. A reasonable choice with regard to the following computations\footnote{For the truncation studied here, a coarse-graining w.r.t. $\Delta_{ab} = -(\Pmu\PMu)_{ab}$, for instance, would not change the discussion of the renormalization in the UV
. In the IR, however, the choice $\tDelta_{ab}$ becomes particularly useful, since it allows for an interesting reformulation of the problem, see Sec. \ref{topoIR}.}
is a coarse-graining with respect to $\tDelta_{ab}$. 
In order to make the computations more transparent, it is furthermore convenient to rescale the regulator and extract a factor\footnote{This rescaling is compatible with the required asymptotic behavior of the regulator owing to the well-established asymptotic freedom of the model with regard to the coupling $g = \zeta^{-1/2}$.} $\zeta_k$. The regulator thus reads
\begin{align}
\label{topoReg}
\mathcal{R}_k = \zeta_k R_k(\tDelta)\quad\quad \Rightarrow \quad\dot{\mathcal{R}}_k = \zeta_k \big(\dot{R}_k(\tilde{\Delta}) - \eta_\zeta R_k(\tilde{\Delta}) \big)~~ \text{with}~~ \eta_\zeta = -\frac{\dot{\zeta}_k}{\zeta_k}\,.
\end{align}
On dimensional grounds $R_k(z)$ has the structure $z~\!\!\!\cdot~\!\!\!r(z/k^2)$. In case a further specification of the regulator is necessary, we will use the ``optimized cutoff'' $R_k(z) = (k^2\!-\!z)\,\Theta(k^2\!-\!z)$ \cite{LitimReg}, with $\Theta(z)$ being the Heaviside step function. \\
The beta functions, $\beta_\zeta \equiv \dot{\zeta}_k$ and $\beta_\theta \equiv \dot{\theta}_k$, can be determined by matching the corresponding operators on both sides of the flow equation. The l.h.s. of Eq. \eqref{erge2} evaluated at $\xi=0$ is simply
\begin{align}
\label{lhs}
 \dot{\Gamma}_k[\varphi]=& \frac{1}{2} \,\beta_\zeta \int d^2 x ~ h_{ab}(\varphi) \pmu \varphi^a \partial^{\mu} \varphi^b + \frac{\ii}{2\pi} \,\beta_{\theta} \int d^2 x \,\emn \sqrt{h} \eab ~\alpha~\pmu \varphi^a \pnu \varphi^b\,.
\end{align}
In order to project the r.h.s. of Eq. \eqref{erge2} onto these operators, an expansion in $B(\alpha)$ is employed which is justified for small fluctuations $\pmu \alpha$ and leads to a separation of symmetric and antisymmetric tensors:
\begin{align}
\label{largeFlow}
 \dot{\Gamma}_k &= \frac{1}{2} \text{Tr}\left\{ \frac{\zeta_k \big(\dot{R}_k(\tilde{\Delta}) - \eta_\zeta R_k(\tilde{\Delta}) \big) }{ \zeta_k R_k(\tilde{\Delta}) + \zeta_k \tilde{\Delta} - B(\alpha) } \right\} \nonumber\\
	&= \frac{1}{2} \text{Tr}\left\{ \frac{\dot{R}_k -\eta_\zeta R_k}{R_{k} +\tilde{\Delta}} ~+~ \zeta^{-1}_k (\dot{R}_k -\eta_\zeta R_k)(R_{k} + \tilde{\Delta})^{-1} B(\alpha) (R_{k} + \tilde{\Delta})^{-1} + O(B^2) \right\} \nonumber\\
	&\equiv \frac{1}{2} \text{Tr}\left\{ \hspace{0.5cm}  W(\tDelta)\hspace{0.5cm}  + ~\zeta^{-1}_k \,B(\alpha)~ f(\tilde{\Delta}) ~+ ~\mathcal{O}(B^2) \right\} \,.
\end{align}
The terms of order $\mathcal{O}(B^2)$ will be neglected in the following analysis. It was explicitly checked that they only yield terms of fourth or higher order in the derivatives which are not considered in the truncation.\\
The first term in \eqref{largeFlow} contains no antisymmetric tensor and hence does not contribute to the running of $\theta$. It will be discussed first. The relevant contributions to $\beta_\theta$ are given by the second term and will be investigated in Sec. \ref{topoUV} and \ref{topoIR}, where it will also become apparent that the second term does not contribute to the running~of~$\zeta$.

\section{The Running of $\zeta$}\label{topoZeta}
The running of $\zeta$ is determined solely by the expression $\tfrac{1}{2}\,\text{Tr}\big\{ W(\tDelta)\big\}\,$, which can be calculated by means of a heat kernel expansion:
\begin{align}
\frac{1}{2}\text{Tr} \big\{ W(\tDelta) \big\} &= ~ \frac{1}{2} \int_0^{\infty} ds~ \tilde{W}(s) ~ \text{Tr}\left\{ \ee^{-s\tilde{\Delta}} \right\} = ~\frac{1}{2} \int_0^{\infty} ds ~\tilde{W}(s) ~\frac{1}{4\pi s} \sum_{n=0}^\infty s^n c_n \,.
\end{align}
The first few coefficients of this heat kernel expansion are well-studied, cf. \cite{Gilkey}. Only $c_1 = -\int_x M^a_{~a}$, with $M_{ab}$ defined in Eq. \!\eqref{variation}, affects the running of $\zeta$, because all coefficients $c_n$ with $n\geq 2$ are of higher orders in the derivatives, and $c_0$ simply yields a field-independent renormalization of the vacuum energy. The $s$-integration for $n=1$ simplifies to $\int_0^{\infty} ds ~\tilde{W}(s)  = W(0)$. For the optimized regulator given above, this expression is equal to $2\!-\!\eta_\zeta$.
The trace of $-M_{ab}$ in target space yields $h_{ab}\,\pmu \varphi^a \pMu \varphi^b$, since $R_{abcd}=h_{ac}h_{bd}-h_{ad}h_{bc}$ on $S^2$, and one can relate both sides of the flow equation such that
\begin{align}
 \frac{1}{2} \, \beta_\zeta \int d^2 x ~& h_{ab}(\varphi) \pmu \varphi^a \partial^{\mu} \varphi^b = \frac{1}{8\pi}(2\!-\!\eta_\zeta)\int d^2 x ~ h_{ab}(\varphi) \pmu \varphi^a \partial^{\mu} \varphi^b \nonumber \\
 \label{topoZetaDot} 
 \Rightarrow &~\beta_\zeta = \frac{1}{4\pi}(2-\eta_\zeta) ~~\Leftrightarrow ~~\beta_\zeta = \frac{2\zeta_k}{4\pi\zeta_k - 1}\,.
\end{align}
Note that $g$ with $\zeta = g^{-2}$ is the usually studied coupling of the model and its beta function is
\begin{equation}
\label{topoGdot}
 \beta_g = -\frac{1}{4\pi}\, g^3 \left( 1- \frac{g^2}{4\pi} \right)^{-1}\,.
\end{equation}
This result confirms the well-known asymptotic freedom of the nonlinear sigma model in two dimensions \cite{Polyakov}. The pole at $g^2= 4\pi$ is only an unphysical artefact\footnote{It is not unusual that such poles occur in FRG computations due to specific features of the regulator without being physically relevant, cf. for instance \cite{codpercrham}.} of the specific regulator choice \eqref{topoReg}. The beta functions \eqref{topoZetaDot} and \eqref{topoGdot} agree with a previous computation within the FRG scheme \cite{PercAle}, apart from an unimportant numerical factor which is due to a slightly different regularization.\\
Since the mass spectrum of the theory, i.e. the threshold in the flow equation, 
depends on $\theta$, one should expect that also $\beta_\zeta$ is affected by this parameter.
The beta function \eqref{topoZetaDot}, however, is independent of $\theta$, and higher orders in $B(\alpha)$ in the expansion \eqref{largeFlow} do not influence the running of $\zeta$, either, but only yield antisymmetric tensors
. The absence of a $\theta$-dependence is not a shortcoming of the specific expansion. In an alternative treatment of the flow equation by means of a heat kernel expansion of a modified Laplacian, which incorporates the derivative operator $B(\alpha)$, $\beta_\zeta$ is also $\theta$-independent.\\ 
A direct investigation of the mass spectrum of the nonlinear sigma model is difficult within the covariant FRG scheme employed here, since the introduction of a mass term for the full field $\phi$ or the background $\varphi$ would break the reparametrization invariance. One could introduce a covariant mass term $m^2_k h_{ab}(\varphi)\xi^a \xi^b$ for the fluctuations and compute its running in the way outlined in \cite{FRGeigen}. However, explicit calculations show that the flow of $m^2_k$ is not affected by $\theta_k$, either. One has to conclude that the chosen ansatz for the effective action is apparently not sensitive to the nontrivial $\theta$-dependence of the spectrum, and one ought to study larger truncations for this purpose.\\

\section{Renormalization of $\theta$ in the UV}
\label{topoUV}
In order to evaluate 
the second term in \eqref{largeFlow}, we can again apply a Laplace transform, $f(\tilde{\Delta}) = \int_0^{\infty} ds\, \tilde{f}(s) \,\exp{(-s\tilde{\Delta})}$ ,
and evaluate the action of $B(\alpha)$ on this expression by means of off-diagonal elements of a heat kernel expansion: 
\begin{align}
\label{ibp}
& \!\!\!\text{Tr} \big\{ \zeta^{-1}_k B(\alpha) f(\tilde{\Delta}) \big\} = \frac{\ii}{\pi} \frac{\theta_k}{\zeta_k} \! \int d^2x \,d^2y\, \emn \sqrt{h} \eab \int_0^{\infty} \!\! ds \,\tilde{f}(s)\!\! \underbrace{\big< x | (\pmu \alpha) \Pnu | y \big>^{bc} }_{=  \pmu \alpha(x) \Pnu (x) \delta(x-y)}\! \underbrace{\big< y| \ee^{-s\tDelta}|x \big>_c^{~a}}_{\equiv \Omega(y,x,s)} \nonumber \\
&= - \frac{\ii}{\pi} \frac{\theta_k}{\zeta_k} \int d^2x\, d^2y\, \emn \sqrt{h} \eab \int_0^{\infty} ds \,\tilde{f}(s) ~ \alpha(x)~ \delta(x-y) \times \nonumber\\
&\hspace{4.7cm}\times\left( \frac{1}{2}H_{\mu\nu}(y) \Omega(y,x,s) + \Pmu(x) \Pnu(y) \Omega(y,x,s) \right)^{ba}\!\!,
\end{align}
where $H_{\mu \nu}$ is the commutator introduces above. Following the reasoning that the limit $\alpha(x)\to 1$ is performed at the end, we can neglect the surface terms coming from integration by parts.  
Since the infinitesimal separation of $x$ and $y$ regularizes the expression and provides access to nontrivial information about the UV, the $\delta$-function $\delta(x-y)$ ought to be understood as limit $y \to x $ which has to be performed carefully.\\
In order to evaluate \eqref{ibp}, appropriate expressions for the off-diagonal elements $\Omega(x,y,s)$ are required. They are derived in App. \ref{HKE} and a computation of their covariant derivatives yields
\begin{align}
\label{topoCalculation}
&\text{Tr} \big\{ \zeta_k^{-1} B(\alpha) f(\tilde{\Delta}) \big\} = - \frac{\ii}{\pi} \frac{\theta_k}{\zeta_k} \int d^2x \, \emn \sqrt{h}\eab ~\alpha(x) \lim_{y\to x} \int_0^{\infty} ds ~\tilde{f}(s)\, \frac{1}{8\pi s^2}\, \ee^{-\frac{|x-y|^2}{4s}} \, \times \nonumber\\
& \hspace{4cm} \times H_{\mu\rho}^{bc}(x)(x-y)^\rho(x-y)_{\nu}~{(c_0)_c}^a(x,y) + \mathcal{O}\big((\partial\varphi)^3\big) \,,
\end{align}
with $c_0(x,y) = \ee^{-\int_y^x \Gamma \partial \varphi\, dx'}$. This exponential function of the pullback connection can be regarded as the identity in the further calculations, as the higher orders in the corresponding series only lead to terms which are beyond the chosen truncation. The tensor $\eab H_{\mu\rho}^{ba}$ is equal to $-2\eab \pmu \varphi^a \partial_\rho \varphi^b$
and the Lorentz indices can be rearranged in two dimensions as follows 
\begin{equation*}
 \eab \emn \pmu \varphi^a \partial_{\rho}\varphi^b (x-y)^{\rho}(x-y)_{\nu} = \frac{1}{2} \eab \emn \partial_{\mu} \varphi^a \partial_{\nu} \varphi^b (x-y)^2\,.
\end{equation*}
The renormalization of the topological parameter $\theta$ can now be determined by a comparison of \eqref{topoCalculation} with the l.h.s. of the flow equation as it is given in Eq. \eqref{lhs}:
\begin{align}
 \frac{\ii}{2\pi}\, \beta_{\theta} \int d^2 x& \,\emn  \sqrt{h} \eab ~\alpha~\pmu \varphi^a \pnu \varphi^b \nonumber \\
&= \frac{\ii}{2\pi}\,\frac{\theta_k}{\zeta_k}  \int d^2x \,\emn \sqrt{h} \eab ~\alpha ~\lim_{y\to x} \int_0^{\infty} ds ~\tilde{f}(s) \frac{(x-y)^2}{8\pi s^2} \ee^{-\frac{|x-y|^2}{4s}}  \pmu \varphi^a \pnu \varphi^b \nonumber \\
\Rightarrow ~\beta_{\theta} = &\frac{\theta_k}{\zeta_k} \,\lim_{u\to 0} \, \int_0^{\infty} ds ~\tilde{f}(s) \, \frac{u^2}{8\pi s^2} ~\ee^{-\frac{u^2}{4s}} \,.
\end{align}
This beta function vanishes for any finite value of $s$ in the limit $u\to 0$. In order to analyze if the limit $s\to 0$ yields relevant contributions, it is useful to notice that the inverse Laplace transform $\tilde{f}(s)$ is in fact a function of $k^2 s$ which can be denoted by $\sigma(k^2s)$:
\begin{align}
\sigma(k^2 s) &=\, \mathcal{L}^{-1}\left[ f(z) \right](s)  
= -\,\mathcal{L}^{-1}\left[k\partial_k\big(z+R_k(z)\big)^{-1}\right](s) - \eta_k\, \mathcal{L}^{-1}\left[\frac{R_k(z)}{\big(z+R_k(z)\big)^2}\right](s) \nonumber \\
 &\equiv\,-k\partial_k\,\sigma_1(k^2 s) - \eta_k\, \sigma_2(k^2 s)\,. 
\end{align}
This can be understood if one considers the Laplace transform at $k=1$ and rescales $z\rightarrow z/k^2$, taking the general structure $z\cdot r(z/k^2)$ of the regulator into account. The case $\sigma_1$, for instance, reads
\begin{align*}
& \big(z+R_{k=1}(z)\big)^{-1} = \int_0^{\infty} ds ~\sigma_1(s)\, \ee^{-sz} \\
&\Rightarrow~ k^2\big(z+R_k(z)\big)^{-1} = \int_0^{\infty} ds ~\sigma_1(s)\, \ee^{-s\tfrac{z}{k^2}} = \int_0^{\infty} ds^\prime  ~ k^2 \sigma_1(k^2s^\prime )\, \ee^{-s^\prime z} \nonumber \, .
\end{align*}
The limit $s\to 0$ can be probed in a controlled way, if one integrates $\beta_\theta = 2\,k^2 \partial_{k^2}\theta$ from the extreme UV down to some finite $k_0$ and applies two substitutions, first $s\rightarrow \tfrac{1}{4}u^2 s$ and then $p^2\equiv \tfrac{1}{4}u^2 k^2 s$:
\begin{align}
\label{limits}
&\theta(\infty) - \theta(k_0^2) =  \int_{k_0^2}^{\infty} dk^2 \lim_{u\to 0} \int_0^{\infty} ds \,\frac{u^2}{8 \pi s^2} \,\ee^{-\frac{u^2}{4s}} \left[- \partial_{k^2} \sigma_1(k^2 s) -\eta_k \frac{1}{2 k^2} \sigma_2(k^2 s)\right] \frac{\theta}{\zeta}(k^2) \nonumber \\
&= \lim_{u\to 0} \, \int_0^{\infty} ds \,\frac{1}{2 \pi s^2}\, \ee^{-\frac{1}{s}} \int_{k_0^2}^{\infty} dk^2\,\left[- \partial_{k^2} \sigma_1\left(\frac{1}{4} u^2 k^2 s\right) -  \eta_k \frac{1}{2 k^2} \sigma_2 \left(\frac{1}{4} u^2 k^2 s \right)\right] \frac{\theta}{\zeta}(k^2)\nonumber \\
&=   \int_0^{\infty} ds \, \frac{1}{2 \pi s^2} \, \ee^{-\frac{1}{s}} \lim_{u\to 0} \int_{\frac{1}{4}k_0^2 u^2 s}^{\infty} dp^2\,\left[- \partial_{p^2} \sigma_1(p^2)~ \frac{\theta}{\zeta}\left(\frac{4 p^2}{u^2 s}\right) \right.\nonumber\\
&\hspace{6.7cm}\left. - \frac{1}{2 p^2} \sigma_2(p^2)~ \eta\left(\frac{4 p^2}{u^2 s}\right)\frac{\theta}{\zeta}\left(\frac{4 p^2}{u^2 s}\right)\right]. 
\end{align}
The limit $u\to 0$ can be performed, while the $s$-integration remains finite and simply yields $\tfrac{1}{2\pi}$. The result is
\begin{align}
\label{UVjump}
\theta(\infty) - \theta(k_0^2)  = \frac{1}{2\pi} \int_{0}^{\infty} dp^2\,\left[ - \partial_{p^2} \sigma_1(p^2) \frac{\theta}{\zeta}(\infty) - \frac{1}{2 p^2} \sigma_2(p^2) \,\eta(\infty)\,\frac{\theta}{\zeta}(\infty)\right].
\end{align}
The $p^2$-integration is finite for an appropriate choice of regulator\footnote{For instance, $\int_0^{\infty} s^{-1} \sigma_2(s) \,ds  =  \int_0^{\infty} dz ~R_{k=1}(z)[z + R_{k=1}(z)]^{-2}$.}. The renormalization of $\theta$ down to any finite scale $k_0$ obviously depends only on the values of $\theta$, $\zeta$ and $\dot{\zeta}$ in the extreme UV and is formally given by a discrete ``jump'' at $k=\infty$. However, it is well-known and was confirmed in Sec. \ref{topoZeta} that the theory is asymptotically free. This statement refers to the coupling $g=\zeta^{-1/2}$, which means that $\zeta$ diverges in the UV. The corresponding beta function, in contrast, remains finite for $\zeta\to\infty$, as given in \eqref{topoZetaDot}. As a result, there is in fact no renormalization of the topological term at any finite $k$, as long as the bare coupling $\theta_\infty$ does not diverge: 
\begin{align}
 \theta_k = \theta_\infty\quad \text{for any}~k>0\,.
\end{align}
This finding agrees with the usual expectation that the topological charge is not renormalized.
However, the argumentation given in Eq. \!\eqref{limits} holds true only for finite $k_0$, but cannot be extended to $k=0$.
A careful investigation of the extreme IR and the zero modes is additionally required and will be given in the following chapter.\\
If one compares the analysis presented above with the one in \cite{ReuterTopo}, the structural similarities between Yang-Mills theory and the nonlinear sigma model are, once more, remarkable. According to \cite{ReuterTopo}, the renormalization of the topological charge in Yang-Mills theories is restricted for $k>0$ to a jump in the extreme UV, similar to \eqref{UVjump}. However, taking the asymptotic freedom of the theory into account (i.e. $\bar{g}\rightarrow 0$) as we did it here, this jump vanishes as well.

\section{Renormalization of $\theta$ in the IR}
\label{topoIR}
In Yang-Mills theory the investigation of the topological parameter in the IR \cite{ReuterTopo} is based on a reformulation of a four-dimensional problem in terms of an eight-dimensional representation of the Clifford algebra \cite{Johansen} which relies on the 't Hooft symbol $\eta_{\alpha\beta\nu}$ \cite{thooft}. A similar reformulation in a ``fermionic language'' is possible in case of the nonlinear sigma model and enables us to study the zero modes. However, since there is no 't Hooft symbol available, we first have to develop a suitable representation of the Clifford algebra.\\
We consider a four-dimensional representation of the gamma matrices $\Gamma_{\mu}$ which is based on two-dimensional matrices $\Omega_{\mu}$ as follows
\begin{align}
 \Gamma_{\mu} \equiv \begin{bmatrix} 0 & \Omega_{\mu} \\ \Omega_{\mu}^\mathsf{T} & 0 \end{bmatrix}\quad\text{with}\quad \Omega_1  \equiv \begin{pmatrix} 0 & 1 \\ -1 & 0 \end{pmatrix} = {\epsilon^a}_b\,,\quad \Omega_2 \equiv \begin{pmatrix} 1 & 0 \\ 0 & 1 \end{pmatrix} = {\delta^a}_b \,.
\end{align}
Note that this construction does not introduce additional spinorial degrees of freedom, but is built upon the symmetric and antisymmetric tensor in the tanget space of the model.
The $\Gamma_\mu$ are defined on the tensor product of the tangent space with itself. 
The identities  
\begin{align}
\label{anticom}
  &\Omega_{\mu} \Omega^\mathsf{T} _{\nu} = \delta_{\mu\nu} \delta^a_{~b} + \epsilon_{\mu\nu} \epsilon^a_{~b}\, \quad\quad \Omega^\mathsf{T} _{\mu} \Omega_{\nu} = \delta_{\mu \nu} {\delta^a}_b - \epsilon_{\mu\nu} \epsilon^a_{~b}\, \nonumber\\
\Rightarrow ~~& \Omega_{\mu}\Omega^\mathsf{T} _{\nu} + \Omega_{\nu}\Omega^\mathsf{T} _{\mu} =\Omega^\mathsf{T} _{\mu}\Omega_{\nu} + \Omega^\mathsf{T} _{\nu}\Omega_{\mu} = 2 \delta_{\mu \nu} {\delta^a}_b\,\quad\quad \Omega_{\mu}\Omega^\mathsf{T} _{\nu} - \Omega^\mathsf{T} _{\mu}\Omega_{\nu} = 2 \epsilon_{\mu\nu} \epsilon^a_{~b}\,
\end{align}
will become useful and ensure the algebraic relation
\begin{align}
\{ \Gamma_\mu , \Gamma_{\nu}\} = \begin{bmatrix} \Omega_{\mu} \Omega^\mathsf{T} _{\nu} + \Omega_{\nu} \Omega^\mathsf{T} _{\mu}& 0\\ 0 & \Omega^\mathsf{T} _{\nu} \Omega_{\mu}  +\Omega^\mathsf{T} _{\nu} \Omega_{\mu} \end{bmatrix} = 2 \delta_{\mu \nu} \id_4\,. 
\end{align}
Moreover, one can define the gamma matrix $\Gamma_*$,
\begin{align}
\Gamma_* = - \begin{bmatrix} \epsilon^a_{~b} & 0 \\ 0& \epsilon^a_{~b} \end{bmatrix} \Gamma_1 \Gamma_2 = \begin{bmatrix} \id_2& 0\\ 0& -\id_2\end{bmatrix},\quad\{ \Gamma_*, \Gamma_{\mu}\} = 0\,,\quad \Gamma_*^2 = \id_4\,,
\end{align}
which provides a notion of chirality. 
In the following computations the Dirac operators 
\begin{equation}
 \slashed{D} \equiv \Gamma_{\mu} \nabla^{\mu},~~ D \equiv \Omega_{\mu} \nabla^{\mu}, ~\text{and}~D^\mathsf{T}  \equiv \Omega^\mathsf{T} _{\mu} \nabla^{\mu}
\end{equation}
will be of particular importance and one may wonder if these expressions are well-defined, since the connection ${\Gamma^a}_{cb}\pmu\varphi^c$ acts on the same space as the gamma matrices. However, both objects are simply linear combinations of $\epsilon^a_{~b}$ and $\delta^a_{~b}$ and hence commute with each other.\\
By means of these Dirac operators the flow equation can be rewritten. According to \eqref{largeFlow}, the running of $\theta$ is determined by:
\begin{align}
\label{FirstOmega}
 \frac{\ii}{2\pi}\, \beta_\theta \int d^2x \, \emn \sqrt{h} \eab&~\alpha ~ \pmu \varphi^a \pnu \varphi^b = \frac{\ii}{2\pi} \frac{\theta_k}{\zeta_k} \int d^2x \,\emn \sqrt{h} \eab~\pmu\alpha(x) \big<x| \Pnu f(\tDelta) |x\big>^{ba}\nonumber\\
=&\, \frac{\ii}{4\pi} \frac{\theta_k}{\zeta_k} \int d^2x ~\pmu\alpha(x) \, \text{tr}_2 \left\{\big<x|(\Omega^{\mu}D^\mathsf{T}  - {\Omega^\mu}^{\mathsf{T}} D) f(\tDelta) |x\big> \right\}\,,
\end{align}
where $\text{tr}_2$ denotes the trace in the two-dimensional tangent space of the model. The flow equation holds true for each field configuration and we can hence evaluate it at a configuration which is convenient from a computational point of view. 
In the present case self-dual fields are a particular useful choice, i.e. fields for which $\pMu \varphi_a = \epsilon^{\mu\rho}\,\eab\, \partial_\rho\varphi^b$. Remembering that $R_{abcd}=h_{ac}h_{bd} - h_{ad}h_{bc}$ and $[\Pmu,\Pnu]_{ab} = R_{abcd}\pmu\varphi^c\pnu\varphi^d$, it is easy to check that for self-dual fields
\begin{align}
 &M_{ab} = \emn \epsilon_{ac}(\Pmu\Pnu)^c_{~b}\,, \nonumber\\
&\tDelta_{ab} = -D^\mathsf{T}  D\,, \quad \tDelta_{ab} -2M_{ab} = -D D^\mathsf{T} \,.
\end{align}
With these relations the r.h.s. of \eqref{FirstOmega} can be written as
\begin{align}
\label{reshape}
&\frac{\ii}{4\pi} \frac{\theta_k}{\zeta_k}\int d^2x ~\pmu\alpha(x) \,\text{tr}_2 \left\{\big<x|(\Omega^{\mu} D^\mathsf{T} - {\Omega^\mu}^{\mathsf{T}} D) f(-D^\mathsf{T}  D) |x\big> \right\}\\
&= \frac{\ii}{4\pi} \frac{\theta_k}{\zeta_k} \int d^2x ~\pmu\alpha(x) \, \text{tr}_2 \left\{\big<x| \Omega^\mu D^\mathsf{T} f(-D D^\mathsf{T} ) - {\Omega^\mu}^{\mathsf{T} } D \,f(-D^\mathsf{T} D) \right. \nonumber\\
&\hspace{6cm}\left.- \Omega^\mu D^\mathsf{T} \big(f(-D D^\mathsf{T} )-f(-D^\mathsf{T}  D)\big)\, |x\big> \right\}\,. \nonumber
\end{align}
The last term in \eqref{reshape} is of order $\mathcal{O}\big((\partial\varphi)^3\big)$ and can be neglected, since $f(-D D^\mathsf{T} )$ and $f(-D^\mathsf{T}  D)$ differ only in terms of second order in the derivatives. 
The two-dimensional trace can now be expressed by means of the gamma matrices as a four-dimensional trace:
\begin{align}
&\frac{\ii}{4\pi} \frac{\theta_k}{\zeta_k} \int d^2x ~\pmu\alpha(x) \, \text{tr}_2 \left\{\big<x| \Omega^\mu D^\mathsf{T}  f(-D D^\mathsf{T} ) - {\Omega^\mu}^{\mathsf{T} }\! D\, f(-D^\mathsf{T}  D) |x\big> \right\} \nonumber \\
&= \frac{\ii}{4\pi} \frac{\theta_k}{\zeta_k} \int d^2x ~\pmu\alpha(x) \, \text{tr}_4 \left\{ \big<x | \begin{bmatrix} 1 & 0 \\ 0& -1 \end{bmatrix} \begin{bmatrix}   \Omega^\mu D^\mathsf{T}  & 0 \\ \!\!0& \!\!{\Omega^\mu}^{\mathsf{T} } D\end{bmatrix} ~f\left( \begin{bmatrix} -D D^\mathsf{T} & 0 \\ \!\!0 & \!\!- D^\mathsf{T}  D\end{bmatrix}\right) | x \big> \right\} \nonumber \\
&= \frac{\ii}{4\pi} \frac{\theta_k}{\zeta_k} \int d^2x ~\pmu\alpha(x) \, \text{tr}_4 \left\{ \big<x |\, \Gamma_* \Gamma^{\mu} \slashed{D} f(-\slashed{D}^2) | x \big> \right\}\,.
\end{align}
In the IR regime the trace is well-defined due to the presence of the regulator and one can integrate by parts\footnote{Assuming appropriate properties of $\alpha(x)$ such that the surface terms can be neglected. Remember that the limit $\alpha(x)\to 1$ is performed at the end.} in order to shift the derivative acting on $\alpha(x)$ to the trace. It acts on bra and ket vector separately and can be contracted\footnote{The matrix $\Gamma^\mu$ anticommutes with $\Gamma_*$ and, utilizing the cyclicality of the trace, it can be contracted with the derivative acting on $|x\big>$. The resulting $\slashed{D}$ then commutes with $\slashed{D}f(-\slashed{D}^2)$.} with $\Gamma^\mu$. The resulting expression shows that only the zero modes provide a non-vanishing contribution:
\begin{align*}
 \frac{\ii}{2\pi}\, \beta_\theta \int d^2x \, \emn \sqrt{h} \eab~\alpha ~ \pmu \varphi^a \pnu \varphi^b =
-\frac{\ii}{2\pi} \frac{\theta_k}{\zeta_k} \int d^2x ~\alpha(x) \, \text{tr}_4 \left\{ \big<x |\, \Gamma_* \slashed{D}^2 f(-\slashed{D}^2) | x \big> \right\}\,.
\end{align*}
The spectrum of $-\slashed{D}^2$ is degenerate and all non-zero-modes appear in pairs of opposite ``chirality'', which cancel each other in the trace due to $\Gamma_*\,$. 
In order to determine the contribution of the zero modes, one can integrate the flow equation between $k=0$ and a finite, but arbitrarily small $k_0$. Since $\dot{\zeta}$ is a continuous function (as confirmed in Sec. \ref{topoZeta}), it is a reasonable approximation to consider $\zeta_k= \zeta_0$ and $\dot{\zeta}_k=\dot{\zeta}_0$ in this infinitesimal momentum range. The renormalization of $\theta$ due to IR effects is hence given as
\begin{align*}
&(\theta_{k_0^2} - \theta_0) \int d^2x \, \emn \sqrt{h} \eab~\alpha ~ \pmu \varphi^a \pnu \varphi^b = \text{Tr}_4 \left\{ \alpha \, \zeta_0^{-1} \Gamma_* \lim_{\lambda\to 0} \int_0^{k_0^2} dk^2~\theta(k^2) \, \lambda f(\lambda) \right\} \nonumber \\
&= \text{Tr}_4 \left\{ \alpha \, \zeta_0^{-1} \Gamma_* \lim_{\lambda\to 0} \int_0^{k_0^2} dk^2~\theta(k^2) \, \lambda \left(-\frac{d}{dk^2}[R_k(\lambda) + \lambda]^{-1} - \frac{1}{2k^2}\eta_{\zeta_0} \frac{R_k(\lambda)}{(R_k(\lambda)+\lambda)^2}\right)\! \right\}\,. 
\end{align*}
Owing to the structure $\lambda \,r(\lambda/k^2)$ of the regulator, one can apply a reparametrization $p^2=\lambda^{-1} k^2$ which yields
\begin{align*}
\text{Tr}_4 \left\{ \alpha \,\zeta_0^{-1} \Gamma_* \lim_{\lambda\to 0} \int_0^{k_0^2/\lambda} dp^2~\theta(p^2\lambda) \, \left(-\frac{d}{dp^2}[R_p(1) + 1]^{-1} - \frac{1}{2p^2}\eta_{\zeta_0}\frac{R_p(1)}{(R_p(1)+ 1)^2}\right) \right\} \,.
\end{align*}
Now the limit $\lambda \to 0$ can be performed. Note that a possible contribution from $p^2 = k_0^2/\lambda \to \infty$ is suppressed by the regulator expressions. The result is
\begin{align*}
(\theta_{k_0^2} - \theta_0) &\int d^2x \, \emn \sqrt{h} \eab~\alpha ~ \pmu \varphi^a \pnu \varphi^b \\
&= - \text{Tr}_4 \left\{ \alpha \,\zeta_0^{-1} \Gamma_* \int_0^{\infty} dp^2~\theta_0 \, \left(\frac{d}{dp^2}[R_p(1) + 1]^{-1} + \frac{\eta_{\zeta_0}}{2p^2}\frac{R_p(1)}{(R_p(1)+ 1)^2}\right) \right\}\, \nonumber
\end{align*}
The first part of the $p$-integral is simply $-\theta_0$ since $\lim_{p\to \infty} R_p(1) = \infty$ and $\lim_{p\to 0} R_p(1) = 0$. In order to compute the second part one has to specify $R_k$. We choose the optimized regulator introduced above, whose rescaled version reads $R_p(1) = (p^2-1)\Theta(p^2-1)$. The integral yields then $\theta_0 \, \tfrac{1}{35} \, \eta_{\zeta_0}$ and we find
\begin{equation}
\label{almostEndTopo}
 (\theta_{k_0^2} - \theta_0) \int d^2x \emn \sqrt{h} \eab~\alpha ~ \pmu \varphi^a \pnu \varphi^b = \frac{\theta_0}{\zeta_0} \left(1-\frac{1}{35}\eta_{\zeta_0}\right) \text{Tr}_4 \big\{ \alpha  \Gamma_* \big\}\,.
\end{equation}
The trace $\text{Tr}_4\left\{ \alpha \Gamma_* \right\}$ ought to be considered in the regularized form $\lim_{s\to 0} \text{Tr}_4\big\{ \alpha \Gamma_* \ee^{s\slashed{D}^2} \big\}$. It represents 
the analytical index of $-\slashed{D}^2$ and can be directly related to the topological index according to the Atiyah-Singer index theorem \cite{Atiyah}. 
An explicit calculation of the trace is given in App. \ref{index} and yields
\begin{equation}
  \lim_{s\rightarrow 0}\text{Tr}_4 \left\{\alpha \, \Gamma_* \, \ee^{s\slashed{D}^2} \right\} = - \frac{1}{2\pi} \int d^2 x \,\emn \sqrt{h} \eab~\alpha~ \pmu \varphi^a \pnu \varphi^b\,.
\end{equation}
The renormalization of $\theta$ in the extreme IR is hence given by
\begin{align}
 \theta_{k_0^2} - \theta_0 = -\frac{1}{2\pi} \frac{\theta_0}{\zeta_0} \left(1-\frac{1}{35}\eta_{\zeta_0}\right)\,.
\end{align}
Since the topological parameter $\theta_{k}$ does not flow from the UV down to any finite scale $k_0$, the relation between bare and full effective coupling is solely determined by this ``jump'' in the IR and reads
\begin{equation}
 \theta_0 = \left(1-\tfrac{1}{2\pi}\big(1-\tfrac{1}{35}\eta_{\zeta_0}\right) \zeta_0^{-1}\big)^{-1} \theta_\infty\,.
\end{equation}
Inserting the result \eqref{topoZetaDot} for $\dot{\zeta}_0$ and rearranging the expression leads to
\begin{equation}
\label{resultTopo}
 \theta_0 = \frac{2\pi \,\zeta_0 \,(4\pi \,\zeta_0 - 1)}{8\pi^2 \,\zeta_0^2 - 6\pi\, \zeta_0  + \frac{33}{35}} ~ \theta_\infty\,.
\end{equation}
The bare and the renormalized parameter are linearly related by a factor that depends only on the effective coupling $\zeta_0$ in the infrared. 
The nonlinear O$(3)$ model apparently constitutes another example of a theory with topological term in which the corresponding parameter is affected by a renormalization in the IR, similar to Yang-Mills and Chern-Simons theory \cite{ReuterTopo,Chern}.
It should be emphasized, yet, that the derivation of \eqref{resultTopo} relied on a generalization of the topological operator by introducing an auxiliary field, for which the limit corresponding to the actual winding number is considered at the end. The physical interpretation of this construction amounts to a topological term which arises from an interaction with a scalar field that assumes a constant expectation value at the end. The finding of this analysis is therefore an interesting, but not decisive statement about the renormalization properties of the nonlinear O$(3)$ model. A next step would be an investigation if a rigorous analysis still shows similar results if one slightly alters the physical interpretation or the corresponding generalization of the topological term.\\
The observed renormalization is an effect of the extreme IR. It thus seems to be impossible to investigate this issue further by means of methods like e.g. lattice computations, which are restricted to finite volumnes. On the other hand, result \eqref{resultTopo} does not contradict recent numerical simulations \cite{Boegli,Nogradi} which showed that the $\theta$-term is a relevant operator and does not renormalize to zero\footnote{Note that the value of the pathologic $\zeta_0 = \tfrac{1}{4\pi}$ is only an artefact of the regulator choice.}.\\
Let us finish with a comment on the periodicity properties: The topological charge is introduced as a phase in the path integral and since the winding number $Q$ assumes integer values for smooth fields, one would expect that the physical properties of the theory are $2\pi$-periodic in $\theta$. The renormalization derived in \eqref{resultTopo}, however, is linear in $\theta$. Although many other analytic and numerical computations, cf. for instance \cite{Asorey,Vicari,DaddaN}, also lack periodicity, it yet demands an explanation. It was conjectured in \cite{DaddaN} that the $|\theta|> \pi$ vacua of the model suffer from a strongly increased pair production which leads to a break down of these vacua until values $|\theta|< \pi$ are reached. This conjecture was motivated by such findings in the massive Schwinger model \cite{Coleman1,Coleman2} which has similar properties as the CP$^n$ models with regard to the vacua properties. In fact, recent large-$n$ computations \cite{Lawrence} indicate that such effects are present in CP$^n$ models as well. Following this argumentation, 
one should trust the result \eqref{resultTopo} only for $\theta < \pi$.

\section{Conclusion}
The renormalization of the topological charge in the $CP^1 \cong \text{O}(3)$ nonlinear sigma model was studied by means of the Functional Renormalization Group. 
A similar approach could be applied as in Yang-Mills theory \cite{ReuterTopo} where a nontrivial renormalization of the topological operator was found in the extreme UV and IR. The approach considers the topological term as the limit of a more general operator in which a space-time dependent topological parameter assumes a constant expectation value.
In order to compute the renormalization in the UV, an off-diagonal heat kernel expansion as well as a careful analysis of a coincidence limit were performed. The extreme IR was studied by means of a reformulation of the flow equation in terms of a specific representation of the Clifford algebra, which enabled to compute the contributions of zero modes using the index theorem.\\
The computations relied on three assumptions: First, the interpretation of the topological term as the limit just mentioned; second, the validity of the chosen regulator of the flow equation; and third, the chosen truncation \eqref{topoAnsatz} of the effective action. The last two assumptions are standard in the FRG framework and this study can hence also be understood as a further test of the method and its applicability to topological aspects. Concerning the chosen truncation, it may be possible that an enlarged truncation could yield more information about the $\theta$-dependence of the mass spectrum which is expected but was not obtained here.\\
The analysis showed that a possible renormalization of $\theta$ in the UV is suppressed by the asymptotic freedom of the model. In the IR, however, a discrete and finite renormalization occurs as an effect of zero modes. In accordance with the findings in Yang-Mills and Chern-Simons theories \cite{ReuterTopo,Chern}, this article thus provides further indications that topological operators can be affected by a renormalization in the extreme IR. An interesting next step would be to study if these indications can be confirmed by an alternative approach which is complementary to the generalization of the topological term that is used in this article. 

\section{Acknowledgments}\label{acknowledgments}

I would like to thank Andreas Wipf, Martin Reuter, Gian Paolo Vacca and Luca Zambelli 
for useful discussions. This work has been supported by the DFG Research Training Group ``Quantum
and Gravitational Fields'' GRK 1523.

\appendix

\section{Off-diagonal Heat Kernel Expansion} 
\label{HKE}
Note that the indices of the target manifold are suppressed for sake of brevity and that the following derivation applies for two dimensions, but could be generalized to other dimensions. Starting with the generic ansatz
\begin{align}
\label{HKEUV}
\Omega(x,y,s) = \big< x \big| \ee^{-s\tDelta} \big| y \big> = \frac{1}{4\pi s} \ee^{-\frac{|x-y|^2}{4s}} \sum_{n=0}^{\infty} s^n c_n(x,y)\,,
\end{align}
the following constraint for $c_n(x,y)$ can be deduced from $\big(\tfrac{d}{ds} +\tilde{\Delta}_x\big)\Omega(x,y,s) = 0$:
\begin{align}
\label{constraintsTopo}
n\, c_n + (x^{\mu} - y^{\mu})\nabla_{x^{\mu}}\, c_n + \tilde{\Delta}_x \, c_{n-1} &= 0.
\end{align}
For $n=0$ the constraint simplifies to $(x^{\mu} - y^{\mu})\nabla_{x^{\mu}} c_0 = 0$ and is solved by
\begin{align}
\label{c0}
 c_0(x,y) = \mathcal{P} ~\ee^{- \int_y^x dz^{\mu}\,\Gamma \pmu \varphi}\,,
\end{align}
where $\mathcal{P}$ denotes the ordering of the operators according to the path from $y$ to $x$, which is understood to be a straight line here. 
The covariant derivative $\nabla_{x^{\mu}} c_0(x,y)$ was discussed in much detail, for instance, in (the appendix of) \cite{Shore} for the case of a gauge field and the result can be transferred to the pullback connection $\Gamma \pmu \varphi$ with little effort. It yields
\begin{align}
\label{derivativeC0}
\nabla_{\!x^{\mu}} \, c_0(x,y) = \int_0^1 dt ~ t\,(x-y)^{\rho} \, c_0(x,z) \, H_{\rho \mu}(z) \, c_0(z,y)\,\quad\text{with}~z= y + t(x-y)\,.
\end{align}
This expression can be expanded in different ways:
\begin{align}
\label{derivativeC0exp}
&\nabla_{\!x^{\mu}} \, c_0 (x,y) \\
&= - \frac{1}{2} c_0(x,y) H_{\mu \rho}(y) (x-y)^{\rho} + \frac{1}{3} c_0(x,y)\nabla_{\sigma} H_{\mu \rho}(y)\! \cdot\! (x-y)^{\sigma} (x-y)^{\rho} + O(x-y)^3 \nonumber \\
&=-\frac{1}{2} H_{\mu \rho}(x) (x-y)^{\rho} c_0(x,y) + \frac{1}{6} \nabla_{\sigma} H_{\mu \rho}(x) \! \cdot\!  (x-y)^{\sigma} (x-y)^{\rho}c_0(x,y) + O(x-y)^3\,. \nonumber
\end{align}
While Eq. \eqref{derivativeC0} proves that $(x^{\mu} - y^{\mu})\nabla_{x^{\mu}} c_0 = 0$ due to the antisymmetry of $H_{\rho \mu}\,$, especially the relations \eqref{derivativeC0exp} will be relevant for the calculation of \eqref{ibp}.
Based on $c_0\,$, a recursive solution for the higher coefficients can be constructed as\footnote{The expression is inspired by the solution to a similar problem in gauge theory \cite{HKELuescher}, which is yet a bit simplier owing to the choice of a specific gauge.}
\begin{align}
\label{topoCoef}
c_n(x,y) = - c_0(x,y)\,\int_0^1 d\lambda ~\lambda^{n-1} \left( c_0^{-1}(x,y)\,\tilde{\Delta}\, c_{n-1}(x,y) \right)^{*\lambda}\,.
\end{align}
The symbol $\big(A(x,y)\big)^{*\lambda}$ denotes an expansion\footnote{Owing to the recursive construction, the coefficient $c_n$ is expandable about $y$ in powers of $(x-y)^{\mu}$ as long as $c_{n-1}$ is, and because $c_0$ is expandable, this holds true for all $c_n$.} of some operator $A(x,y)$ about $y$ in powers of $(x - y)^{\mu}$, in which each factor $(x - y)^{\mu}$ is multiplied by $\lambda$. Although this expression is rather abstract, it will be sufficient for the purposes of this investigation. Remembering that $(x^{\mu} - y^{\mu})\nabla_{\!x^{\mu}}\, c_0 = 0$, it indeed provides the correct off-diagonal heat kernel coefficients:
\begin{align}
& (x^{\mu}\! -\! y^{\mu})\nabla_{x^{\mu}}\, c_n \\
&= - c_0 \int_0^1 d\lambda ~\lambda^{n-1} (x-y)^{\mu} \partial_{x^{\mu}} \left( c_0^{-1} \tilde{\Delta}\, c_{n-1} \right)^{*\lambda}
=- c_0 \int_0^1 d\lambda ~\lambda^{n-1} \lambda \frac{\partial}{\partial \lambda} \left( c_0^{-1} \tilde{\Delta}\, c_{n-1} \right)^{*\lambda} \nonumber \\
&= - c_0 \left[ \lambda^n \left( c_0^{-1} \tilde{\Delta}\, c_{n-1} \right)^{*\lambda} \right]^{\lambda = 1}_{\lambda=0} + n\, c_0 \int_0^1 d\lambda ~\lambda^{n-1} \left( c_0^{-1} \tilde{\Delta}\, c_{n-1} \right)^{*\lambda} 
= - \tilde{\Delta}\, c_{n-1} - n\, c_n \nonumber
\end{align} 
Based on this expansion of $\Omega(x,y,s)$ one can evaluate the trace \eqref{ibp}. According to \eqref{topoCoef}, all $c_n$ with $n\geq1$ are of second or higher order in the derivatives, such that the action of $\Pmu(y) \Pnu(x)$ on these coefficients yields only terms of fourth or higher order in the derivatives which are not considered in our truncation. The derivatives of $c_0$ are given in \eqref{derivativeC0exp}.

\section{Index of Dirac Operator}
\label{index}
In order to compute $\lim\limits_{s\to 0} \text{Tr}_4\big\{ \alpha \Gamma_* \ee^{s\slashed{D}^2} \big\}$ one can employ a heat kernel expansion similiar to Eq. \!\eqref{HKEUV}. Starting with the ansatz
\begin{equation}
 \big< x| \ee^{s\slashed{D}^2} | y \big> = \frac{1}{4\pi s} \,\ee^{-\tfrac{|x-y|^2}{4s}} \sum_{n=0}^{\infty} s^n C_n (x,y)\,,
\end{equation}
where $C_n$ are $4\times4$ matrices defined on the tensor product of the target space with itself, constraints for these coefficients can be derived in the same way as in Eq. \!\eqref{constraintsTopo} and read:
\begin{align*}
n C_n \!+ (x-y)^{\mu} \begin{bmatrix} \Pmu & \!\!\!\!0\\ 0 & \!\!\!\!\Pmu \end{bmatrix}\! C_n -\!\! \begin{bmatrix} \nabla^{\mu} \Pmu + {\epsilon}_{ab} \emn \Pmu \Pnu & \!\!0\\ 0 & \!\!\nabla^{\mu} \Pmu - {\epsilon}_{ab} \emn \Pmu \Pnu \end{bmatrix}\! C_{n-1} = 0\,.
\end{align*}
The relevant contribution to the index is provided by $C_1$, since all higher coefficients are suppressed in the limit $s\to 0$, while $C_0$ only yields a field-independent vacuum renormalization. The coefficient $C_1$ can be constructed from the solution
\begin{align}
 C_0 = \begin{bmatrix} c_0 & 0 \\ 0& c_0 \end{bmatrix}\,,~\text{with}~c_0~\text{given in Eq. \!\eqref{c0}}\,,
\end{align}
analogously to \eqref{topoCoef} as 
\begin{align}
 C_1 = \begin{bmatrix} c_1^+ & 0\\ 0& c_1^-\end{bmatrix}~~\text{with}~~&c_1^+ = c_0 \int_0^1 d\lambda~ \big( c_0^{-1} \, (\nabla^{\mu}\Pmu + {\epsilon}_{ab} \emn \Pmu \Pnu ) \, c_0 \big)^{*\lambda}\\
&c_1^- = c_0 \int_0^1 d\lambda ~\big( c_0^{-1}\, (\nabla^{\mu}\Pmu - {\epsilon}_{ab} \emn \Pmu \Pnu )\, c_0 \big)^{*\lambda}\,.
\end{align}
Multiplying $C_1$ by $\Gamma_*$ and taking the trace, the terms containing $\nabla^{\mu}\Pmu$ cancel each other, while the terms containing ${\epsilon}_{ab} \emn \Pmu \Pnu$ add up. Moreover, we know that 
\begin{equation}
 {\epsilon}_{ac}\emn {(\Pmu \Pnu)^c}_b = \frac{1}{2} \epsilon_{ac} \emn R^c_{~bde} \pmu \varphi^d \pnu \varphi^e = \epsilon_{ad} \emn \pmu \varphi^d \pnu \varphi_b \,.
\end{equation}
Finally, the coincidence limit $ y \to x$ is taken such that $c_0 \to \id_2$ and the trace yields
\begin{align}
 \lim_{s\rightarrow 0}\text{Tr}_4 \left\{\alpha \Gamma_* \ee^{s\slashed{D}^2} \right\} = - \frac{1}{2\pi} \int d^2 x \, \emn \sqrt{h} \eab~\alpha~ \pmu \varphi^a \pnu \varphi^b\,.
\end{align}

\bibliographystyle{model1a-num-names.bst}
\bibliography{FRGtopologicalNLSM}

\end{document}